\documentclass[usenatbib]{mn2e}
\usepackage{graphicx,times}
\usepackage{bm}
\usepackage{epsf}

\usepackage{wick}

\def\qtwo{\qquad\qquad}

\newcommand{\n}{\noindent}

\date{\today,~ $ $Revision: 0.9 $ $}

\begin{document}

\onecolumn

\title[A New Approach to Probing Primordial Non-Gaussianity]
{A New Approach to Probing Primordial Non-Gaussianity}

\author[Munshi \& Heavens]
{Dipak Munshi and Alan Heavens \\
Scottish Universities Physics Alliance (SUPA), Institute for Astronomy, University of Edinburgh, Blackford Hill,  Edinburgh EH9 3HJ, UK }

\maketitle

\begin{abstract}
We address the dual challenge of estimating deviations from Gaussianity arising in models of the Early Universe, whilst retaining information necessary 
to assess whether a detection of non-Gaussianity is primordial. We do this by constructing a new statistic, the bispectrum-related power spectrum, which is constructed from a map of the Cosmic Microwave Background. The estimator is optimised for primordial non-Gaussianity detection, but can also be useful in distinguishing primordial non-Gaussianity from secondary non-Gaussianity, such as may arise from unsubtracted point sources, or residuals from component separation.  Extending earlier studies we present unbiased non-Gaussianity estimators optimised for partial sky coverage and inhomogeneous noise associated with realistic scan strategies, but which retain the ability to assess foreground contamination.
\end{abstract}

\begin{keywords}
Cosmology: theory -- cosmic microwave-background -- large-scale structure of Universe
Methods: analytical -- Methods: statistical --Methods: numerical
\end{keywords}

\section{Introduction}
\label{sec:intro}

The statistical properties of fluctuations in the cosmic microwave background (CMB) radiation can be used to probe the very
earliest stages of the Universe's history, and provide valuable information on the mechanisms
which ultimately gave rise to the existence of structure within the Universe. This may include
evidence for inflation, the process by which the rapid expansion of the Universe is thought to
have arisen. In standard inflationary models, the fields in the early Universe should be very
close to random Gaussian fields, so a detection of large non-Gaussianity would be highly
significant, and may indicate a different history, such as warm inflation or multiple-field
inflation, or a completely different mechanism such as those arising from topological defects.
A difficulty for methods designed to detect non-Gaussianity in the CMB is that other processes can contribute, such as gravitational lensing,
unsubtracted point sources, and imperfect subtraction of galactic foreground emission (e.g. \citet{GoldbergSpergel99,CoorayHu,VerdeSperg02,Castro04,Babich08}).
The challenge therefore is to provide evidence that any detection of non-Gaussianity is primordial
in origin, and not a result of these other effects.
The aim of this paper is to provide an optimised framework not just for detecting non-Gaussianity, but for
assessing the contributions from various sources.

Non-Gaussianity from simplest inflationary models based on a single slowly-rolling scalar field is
typically very small \citep{Salopek90,Salopek91,Falk93,Gangui94,Acq03,Mal03},  (see \citet{Bartolo06}
and references there in for more details). Variants of simple inflationary models such as multiple
scalar fields \citep{Lyth03}, features in the inflationary potential, non-adiabatic fluctuations, non-standard
kinetic terms, warm inflation \citep{GuBeHea02,Moss}, or
deviations from Bunch-Davies vacuum can all lead to much higher level of non-Gaussianity.  Early observational work on the bispectrum
from COBE \citep{Komatsu02} and MAXIMA \citep{Santos} was followed by much more accurate analysis with WMAP \citep{Komatsu03,Crem07a,Spergel07}.
With the recent claim of a detection of non-Gaussianity \citep{YaWa08} in the Wilkinson
Microwave Anisotropy Probe 5-year (WMAP5) sky maps, interest in non-Gaussianity has obtained a
tremendous boost. Much of the interest in primordial non-Gaussianity has focussed on a
phenomenological `{\em local} $f_{NL}$' parametrisation in terms of the perturbative non-linear
coupling in the primordial curvature perturbation \citep{KomSpe01}:
\begin{equation}
\Phi(x) = \Phi_L(x) + f_{NL} ( \Phi^2_L(x) - \langle \Phi^2_L(x) \rangle ),
\end{equation}
where $\Phi_L(x)$ denotes the linear Gaussian part of the Bardeen curvature and $f_{NL}$ is the
non-linear coupling parameter. A number of models have non-Gaussianity which can be
approximated by this form. The leading order non-Gaussianity therefore is at the level of the
bispectrum, or in configuration space at the three-point level.
Many studies involving primordial non-Gaussianity have used the bispectrum, motivated by the
fact that it contains all the information about $f_{NL}$ \citep{Babich}. It has been extensively
studied \citep{KSW,Crem03,Crem06,MedeirosContaldi06,Cabella06,Liguori07,SmSeZa09}, with most of these measurements providing 
convolved estimates of the bispectrum. Optimised 3-point
estimators were introduced by \citet{Heav98}, and have been successively developed
\citep{KSW,Crem06,Crem07b,SmZaDo00,SmZa06} to the point where an estimator for $f_{NL}$ which saturates
the Cramer-Rao bound exists for partial sky coverage and inhomogeneous noise \citep{SmSeZa09}.
Approximate forms also exist for {\em equilateral} non-Gaussianity, which may arise in models
with non-minimal Lagrangian with higher-derivative terms \citep{Chen06,Chen07}. In these models, the largest
signal comes from spherical harmonic modes with $\ell_1\simeq \ell_2 \simeq \ell_3$, whereas for
the local model, the signal is highest when one $\ell$ is much smaller than the other two --
the so-called {\em squeezed} configuration.

Reducing the CMB data to a single lossless estimate for $f_{NL}$ is extremely elegant, but it
suffers from the disadvantage that a single number loses the ability to determine the extent to
which the estimate has been contaminated by non-primordial signals. What we seek to
to perform a less aggressive data compression of the CMB data, not to a single number, but to a
function, which has known expected form for primordial models, and for which the contributions
from other signals can be estimated. The purpose of this is to be able to demonstrate that a
non-Gaussian signal is indeed primordial, or alternatively accounted for by non-primordial
signals. We do this in a way which is still optimal for local or equilateral $f_{NL}$ models, although the formalism is general. The function we
choose is the integrated cross-power spectrum of pair of maps constructed from the CMB data.
Mathematically, it is closely related to previous estimators, but the interpretation of the
output is different, and offers very significant advantages.

This paper is organised as follows: section \$\ref{sec:infl_models} provides a small review of
available models of primordial non-Gaussianity.
Section \$\ref{sec:cmb_bispec} relates the projected-bispectrum to the corresponding primordial bispectrum.
Section \$\ref{sec:all_sky} presents the optimised bispectrum-related power spectrum estimator for 
the idealised case of an all-sky survey with
homogeneous noise. This is not optimal for partial sky coverage or inhomogeneous noise, but is
straightforward and shows the connection with the $f_{NL}$ estimator of \cite{KSW}. Here we
present the theoretical expectation for the local $f_{NL}$ model, and show the link between the
$f_{NL}$ estimator based
on one-point statistics and its two-point counterpart. 
Section \$\ref{sec:part_sky} provides a method which can handle partial sky coverage
and non-uniform noise in an approximate way using Monte-Carlo simulations.   
In section \$\ref{sec:generalcase} we provide the full optimal weights for bispectrum analysis in
the presence of sky-cuts and the inhomogeneous noise.
The full inverse-covariance of the data is introduced which makes the
estimator an optimal one. The final section is devoted to discussion
and future plans for numerical implementation.

\section{Models of Primordial non-Gaussianity}
\label{sec:infl_models}

Deviations from pure Gaussian statistics can provide direct clues regarding
inflationary dynamics. The single-field slow-roll model of inflation provides a very small
level of departure from Gaussianity, far below present experimental detection
limits \citep{Mal03,Acq03}. Many other variants however will produce
a much higher-level of non-Gaussianity which will be within the reach of all-sky
experiments such as Planck. 

In general models can be distinguished by the way they predict coupling between different Newtonian potential modes:
\begin{equation}
\langle \phi({\bf k}_1)\phi({\bf k}_2)\phi({\bf k}_3)\rangle = (2\pi)^3 \delta^{3 \rm D}({\bf k}_1 + {\bf k}_2 + {\bf k}_3)F(k_1,k_2,k_3).
\end{equation}
\n
The function F encodes the information about mode-mode coupling and $\delta^{3\rm D}({\bf k}_1 + {\bf k}_2 + {\bf k}_3)$
ensures triangular equality in $k$-space. Different models of inflation are typically
divided into two different groups. The first class of model is known as the ``local model'', where the contribution from $F(k_1,k_2,k_3)$ is largest when the 
wavevectors are in the so-called ``squeezed'' configuration, where e.g. $k_1 \ll k_2,k_3$. The local form of non-Gaussianity is predominant
in  models where there is non-linear coupling between the field driving inflation (the inflaton)
and the curvature perturbations \citep{Salopek90,Salopek91,Gangui94}, such as the curvaton
model \citep{Lyth03} and the Ekpyrotic model \citep{Koya07,Buch08}. The primordial bispectrum in the local model can be written as:
\begin{equation}
F^{loc}(k_1,k_2,k_3) = f_{NL}^{loc} \left [ {\Delta^2_{\phi} \over k_1^3 k_2^3} + {\rm cyc.perm.} \right ] ,
\end{equation}
\n
where the power spectrum of inflationary curvature perturbations is given by $P_{\Phi} = \Delta_{\Phi} / k^3 $, in general for deviation from Harrison-Zeldovich power-spectra one has $P_{\Phi} = \Delta_{\Phi} / k^{4-n_s} $.

The other main class consists of models where the contribution from $F(k_1,k_2,k_3)$ is maximum
for configurations where $k_1 \sim k_2 \sim k_3 $. The equilateral form appears from non-canonical kinetic  
terms such as Dirac-Born-Infield (DBI) action (Alishahiha et. al. 2004), the ghost condensation
(Akrani-Hamed et al. 2004) or various single-field models where the scalar field acquires a 
low sound speed \citep{Chen07,Cheung08}.  The equilateral model is not a separable function of the $k_i$, which complicates the analysis considerably, but it was shown by \citet{Crem06} that the following approximate 
form can model the equilateral case very accurately:
\begin{equation}
F^{eq}(k_1,k_2,k_3) = f_{NL}^{eq} \left [ -3 { \Delta_{\phi}^2 \over k_1^3 k_2^3 }
 -2  { \Delta_{\phi}^2 \over k_1^2 k_2^2 k_3^2 } + 6  { \Delta_{\phi}^2 \over k_1 k_2^2 k_3^3 }+ {\rm cyc.perm.} \right ] .
\end{equation}
\n
Secondary non-Gaussianity resulting from various sources e.g. coupling of lensing and the Sunyaev-Zel'dovich effect, or lensing
and the Integrated Sachs-Wolfe effect can also contribute to the observed bispectrum \citep{SperGold99a,SperGold99b}. We will present a general analysis of secondary
bispectra as well as the one induced by quasi-linear evolution of gravitational perturbations 
\citep{MuSoSt95}.

\section{Angular CMB Bispectrum with Primordial Non-Gaussianity}
\label{sec:cmb_bispec}

The angular bispectrum can be defined as the three-point correlation in the
harmonic domain.  With temperature fluctuations as a function of solid angle $\hat\Omega$, $\Delta T(\hat\Omega)$,
\begin{equation}
a_{lm} \equiv \int d\hat\Omega \frac{\Delta T(\hat\Omega)}{T} Y_{lm}^*(\hat\Omega)
\end{equation}
and the three-point function may be written
\begin{eqnarray}
\langle a_{lm}a_{l'm'}a_{l''m''} \rangle
\equiv B_{ll'l''} \left ( \begin{array}{ c c c }
     l & l' & l'' \\
     m & m' & m''
  \end{array} \right).
\end{eqnarray}
\n
This form preserves the the rotational invariance of the three-point correlation function in the harmonic domain. 
The quantity in parentheses is the Wigner 3j-symbol, which is nonzero only for triplets ($l_1,l_2,l_3$) which satisfy the triangle rule, including
that the sum $l_1+l_2+l_3$ is even,  ensuring the parity invariance of the bispectrum.
The reduced bispectrum $b_{ll'l''}$ was introduced by \citet{KomSpe01} which will be helpful (see \citet{BCP04} for elaborate discussion):
\begin{eqnarray}
B_{ll'l''} \equiv  \sqrt {(2l+1)(2l'+1)(2l'+1)\over 4\pi}\left ( \begin{array}{ c c c }
     l & l' & l'' \\
     0 & 0 & 0
  \end{array} \right)b_{ll'l''} \equiv I_{ll'l''}b_{ll'l''}.
\end{eqnarray}
\n
The reduced bispectrum $b_{ll'l''}$ can be expressed in terms of the kernel $F(k_1,k_2,k_3)$ for various
models that we will be considering:
\begin{equation}
b_{l_1l_2l_3} =  { \left ( 2 \over \pi \right ) }^3 \int dr r^2 
\int k_1^2 dk_1 j_{l_1}(k_1r)\Delta_{l_1}^T(k_1r)
\int k_2^2 dk_2 j_{l_2}(k_2r) \Delta_{l_2}^T(k_2r)
\int k_3^2 dk_3 j_{l_3}(k_3r)\Delta_{l_3}^T(k_3r) F(k_1,k_2,k_3) ;
\end{equation}
\n
where $\Delta^T_l(k)$ denotes the transfer function which relates the inflationary potential $\Phi$ to
the spherical harmonics $a_{lm}$ of the temperature perturbation in the sky (e.g. \citet{WangKam00}):
\begin{equation}
a_{lm} = 4\pi (-i)^l \int { d^3k \over (2\pi)^3 } \Phi({\bf k}) \Delta_l^T(k) Y_{lm}^*(\hat {\bf k}).
\end{equation}
\n
Using these definitions one can express the reduced bispectra for the local
and equilateral case as follows:
\begin{eqnarray}
&&b_{l_1l_2l_3} = 2 f_{NL}^{loc} \int r^2 dr \left [ \alpha_{l_1}(r) \beta_{l_2}(r) \beta_{l_3}(r) + {\rm cyc.perm.} \right ] \\
&&b_{l_1l_2l_3} = 6 f_{NL}^{eq}\int r^2 dr \left [-\alpha_{l_1}(r) \beta_{l_2}(r) \beta_{l_3}(r) - 2\delta_{l_1}(r) \delta_{l_2}(r) \delta_{l_3}(r) + \beta_{l_1}(r)\gamma_{l_2}(r) \delta_{l_3}(r) + {\rm cyc.perm.} \right ] 
\end{eqnarray}
\n
We will use these forms to construct associated fields $A,B$ etc. from
temperature fields with appropriate weighting to optimise our estimator.
We list the explicit expressions for the functions $\alpha_l(r),\beta_l(r)$ etc.
for completeness \citep{Crem06}:
\begin{eqnarray}
\alpha_l(r) &\equiv& {2 \over \pi} \int_0^{\infty} k^2 dk \Delta_l(k) j_l(kr)  \nonumber\\
\beta_l(r) &\equiv &  {2 \over \pi} \int_0^{\infty} k^2 dk P_{\Phi}(k)\Delta_l(k) j_l(kr) \nonumber\\
\gamma_l(r) &\equiv& {2 \over \pi} \int_0^{\infty} k^2 dk P_{\Phi}^{1/3}(k) \Delta_l(k) j_l(kr)  \nonumber\\
\delta_l(r) &\equiv& {2 \over \pi} \int_0^{\infty} k^2 dk P_{\Phi}^{2/3}(k)\Delta_l(k) j_l(kr) 
\end{eqnarray}
\n
Numerical evaluations of these functions can be performed by using the publicly available
software such as CAMB or CMBFAST.

\section{All Sky analysis with homogeneous noise}
\label{sec:all_sky}

In this section, we compute the main statistic which will be used to estimate primordial non-gaussianity,
and which can also be used to assess whether a non-gaussian signal is indeed primordial.  We call the statistic the {\em bispectrum-related power spectrum}, as it derives from the cross-power spectrum of certain maps constructed from the CMB map data, and which, as we will see, is related to the primordial nongaussianity.

The analysis in this section is optimal for detecting primordial non-gaussianity in the case of all-sky coverage and homogeneous noise.  These assumptions will not hold in practice, but we present this simpler case for clarity first, and to show the connection with previous work.  We relax the assumptions later, and give optimised estimators for realistic cases in later sections.

\subsection{Local Model}

Following \citet{KSW}, we first construct the 3D 
fields $A(r,\hat\Omega)$ and $B(r,\hat\Omega)$ from the expansion coefficients of the observed CMB map, $a_{lm}$.  The harmonics here $A_{lm}(r)$ and $B_{lm}(r)$ are simply weighted
spherical harmonics of the temperature field $a_{lm}$ with weights constructed
from the CMB power spectrum $C_l$ and the functions $\alpha_l(r)$ and $\beta_l(r)$ respectively:
\begin{eqnarray}
A(r,\hat \Omega) \equiv \sum_{lm} Y_{lm}(\hat \Omega) A_{lm}(r); ~~~ A_{lm}(r) \equiv {\alpha_{l}(r) \over C_l} b_l a_{lm} \\
B(r,\hat \Omega) \equiv \sum_{lm} Y_{lm}(\hat \Omega) B_{lm}(r); ~~~ B_{lm}(r) \equiv {\beta_{l}(r) \over C_l} b_l a_{lm}.
\end{eqnarray}
\n
The function $b_l$ represents beam smoothing, and from here onward we will absorb it into 
the harmonic transforms.   Using these definitions \citet{KSW} define the one-point mixed-skewness for the fields 
$A(r,\hat \Omega)$ and $B(r,\hat \Omega)$:
\begin{equation}
S^{loc}_3 = S^{AB^2}_3 \equiv \int r^2 dr \int d \hat \Omega  A(r,\hat \Omega )B^2(r,\hat \Omega). 
\end{equation}
\n
$S_3$ can be used to estimate $f^{loc}_{NL}$, but such radical data compression to a single number loses the ability to estimate contamination of the estimator by other sources of nongaussianity.  As a consequence, we construct a less radical compression, to a function of $l$ which can be used to estimate $f_{NL}^{loc}$, but which can also be analysed for contamination by, for example, foregrounds.  We do this by constructing the integrated cross-power spectrum of the maps $A(r,\hat\Omega)$ and $B^2(r,\hat\Omega)$.  Expanding $B^2$ in spherical harmonics gives
\begin{eqnarray}
B_{lm}^{(2)}(r) &\equiv& \int d\hat \Omega B^2(r,\hat \Omega)Y_{lm}(\hat \Omega) \nonumber \\
&=& \sum_{l'm'} \sum_{l''m''} {\beta_{l'}(r) \over C_{l'}}{\beta_{l''}(r) \over C_{l''}}
 \sqrt {(2l+1)(2l'+1)(2l''+1) \over 4\pi}\left ( \begin{array}{ c c c }
     l & l' & l'' \\
     0 & 0 & 0
  \end{array} \right)
\left ( \begin{array}{ c c c }
     l & l' & l'' \\
     m & m' & m''
  \end{array} \right)a_{l'm'} a_{l''m''}
\end{eqnarray}
\n
and we define the cross-power spectrum $C_l^{A,B^2}(r)$ at a radial
distance $r$ as
\begin{equation}
C_l^{A,B^2}(r) = {1 \over 2l + 1} \sum_m {\rm Real}\left \{ A_{lm}(r) B_{lm}^{(2)}(r) \right \},
\end{equation}
Integrating this over $r$ gives:
\begin{equation}
C_l^{A,B^2}\equiv \int r^2 dr~ C_l^{A,B^2}(r).
\end{equation}
\n
This integrated cross power spectrum of $B^2(r,\hat\Omega)$ and $A(r,\hat\Omega)$ carries information about the underlying
bispectrum $B_{ll'l''}$, as follows:
\begin{eqnarray}
C_l^{A,B^2} =  {1 \over 2l + 1} \sum_m \sum_{l'm'}\sum_{l''m''}&& \int r^2 dr \left \{ {\alpha_{l}(r) \over C_l}{\beta_{l'}(r) \over C_{l'}}{\beta_{l''}(r) \over C_{l''}} \right \}a_{lm}a_{l'm'}a_{l''m''} \nonumber\\
&& \times \sqrt {(2l+1)(2l'+1)(2l''+1) \over 4\pi}\left ( \begin{array}{ c c c }
     l & l' & l'' \\
     0 & 0 & 0
  \end{array} \right)
\left ( \begin{array}{ c c c }
     l & l' & l'' \\
     m & m' & m''
  \end{array} \right). 
\end{eqnarray}
\n
Similarly we can construct the cross power spectrum of the product map $AB(r,\hat\Omega)$ and 
$B(r,\hat\Omega)$, which we denote as $C_l^{B,AB}$;
\begin{eqnarray}
C_l^{AB,B} =  {1 \over 2l + 1} \sum_m \sum_{l'm'}\sum_{l''m''}&& \int r^2 dr \left \{ {\beta_{l}(r) \over C_l}{\alpha_{l'}(r) \over C_{l'}}{\beta_{l''}(r) \over C_{l''}}  \right \}a_{lm}a_{l'm'}a_{l''m''} \nonumber\\
&& \times \sqrt {(2l+1)(2l'+1)(2l''+1) \over 4\pi}\left ( \begin{array}{ c c c }
     l & l' & l'' \\
     0 & 0 & 0
  \end{array} \right)
\left ( \begin{array}{ c c c }
     l & l' & l'' \\
     m & m' & m''
  \end{array} \right). 
\end{eqnarray}
\n
Using these expressions, and the following relation, we can write this more compactly in terms of 
the estimated CMB bispectrum
\begin{equation}
\hat B_{ll'l''} = \sum_{mm'm''} \left ( \begin{array}{ c c c }
     l & l' & l'' \\
     m & m' & m''
  \end{array} \right) a_{lm}a_{l'm'}a_{l''m''}
\end{equation}
\n
from which we compute our new statistic, the {\em  bispectrum-related power spectrum}, $C_l^{loc}$ as
\begin{equation}
C_l^{loc} \equiv (C_l^{A,B^2} + 2 C_l^{AB,B}) = {\hat f^{loc}_{NL} \over (2l+1)} \sum_{l'}\sum_{l''} \left \{ {B^{loc}_{ll'l''} \hat B_{ll'l''} \over C_l C_{l'} C_{l''}} \right \} 
\label{BiPS}
\end{equation}
\n
where $B^{loc}_{ll'l''}$ is the bispectrum for the local $f_{NL}$ model, normalised to $f_{NL}^{loc}=1$.  We can now use standard statistical techniques to estimate $f_{NL}^{loc}$. Note that if we sum over all $l$ values then we recover the estimator $S_{prim}$ 
of \citet{KSW}, which is the cross-skewness of $ABB$:

\begin{equation}
S^{loc}_3 \equiv S^{AB^2}_3 = \sum_l (2l+1) ( C_l^{A,B^2} + 2 C_l^{AB,B} ) = \hat f_{NL}^{loc} \sum_l \sum_{l'}\sum_{l''} \left \{ {B_{ll'l''}^{loc}\hat B_{ll'l''} \over C_l C_{l'} C_{l''}} \right \}.
\end{equation}

\begin{figure}
\begin{center}
{\epsfxsize=6. cm \epsfysize=6. cm {\epsfbox[30 146 590 854]{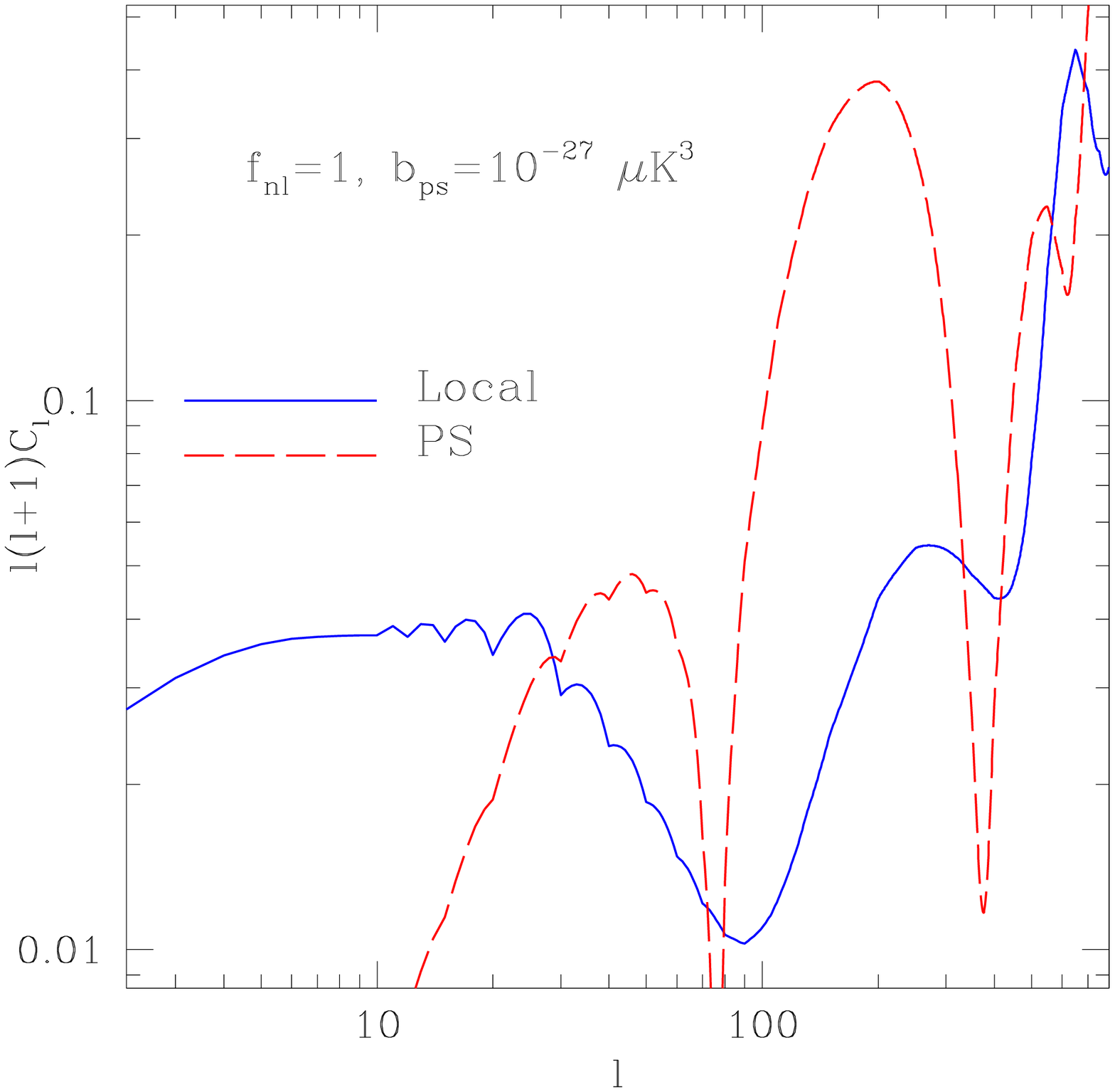}}}
\end{center}
\caption{The bispectrum-related power spectrum is plotted as function
of angular scale $l$. The red curve corrspond to the point source
cross-contamination $b_{ps}=10^{-27}\mu K$ and the bllue curve correspond to local model with $f_{NL}=1$. See text for details.}
\label{fig:bicls}
\end{figure}
\n
We show in Fig. 1 the form of the bispectrum-related power spectrum for the local model.   If the signal observed is inconsistent with this form, it would indicate a departure from this form of primordial non-Gaussianity, and/or a significant contamination by foreground sources.   In the right hand panel we show the expected form of contamination of the $C_l^{loc}$ statistic by a simple foreground source: unsubtracted point sources, randomly distributed.  The contamination scales with the number density.  This contamination is expected to be very low \citep{KomSpe01}, but we show it as an illustration of how foreground effects may be detected.  Since its structure is very different from the local primordial signal, it should be relatively easy to decouple.  Note that the point source contamination here is the contribution to the local model bispectrum-related power spectrum - i.e. it is equation (\ref{BiPS}) with one local $B$ and one point source $B$ (constant $b_{ll'l''}$), not two point source $B$ terms.

\subsection{Equilateral Model}

\n
The form for the reduced bispectrum for the equilateral model as mentioned above is an
approximation to the
real bispectrum generated in theories with higher-order derivatives in the Lagrangian.
In addition to the quantities $A$ and $B$ we have constructed analogous fields $C$ and $D$
with corresponding weights $\gamma_l(r)$ and $\delta_l(r)$:
\begin{eqnarray}
C(r,\hat \Omega) \equiv \sum_{lm} Y_{lm}(\hat \Omega) C_{lm}(r); ~~~ C_{lm}(r) \equiv {\gamma_{l}(r) \over C_l} b_l a_{lm} \\
D(r,\hat \Omega) \equiv \sum_{lm} Y_{lm}(\hat \Omega) D_{lm}(r); ~~~ D_{lm}(r) \equiv {\delta_{l}(r) \over C_l} b_l a_{lm}.
\end{eqnarray}
\n
From these fields and using $A$ and $B$ previously defined a one-point statistic
can be constructed \citep{Crem06}:
\begin{equation}
S^{eq}_3 = -18 \{ S^{AB^2}_3-{2\over 3}S^{D^3}_3-2S_3^{BCD} \}  = -{18}\int r^2 dr \int d \hat \Omega \left [ A(r,\hat \Omega)B(r,\hat \Omega)^2 + {2 \over 3}D(r,\hat \Omega )^3 
- 2B(r,\hat \Omega )C(r,\hat \Omega)D(r,\hat \Omega) \right ].
\end{equation}
\n
The associated power spectrum will have a composite structure with many contributing terms.
\begin{equation}
C_l^{eq} = -18 \left \{ (C_l^{A,B^2}+2C_l^{AB,B}) -2C_l^{D,D^2} -2 (C_l^{B,CD}+C_l^{C,BD}+C_l^{D,BC}) \right \}
\end{equation}
\n
Following the same procedure outlined above one can now write:
\begin{equation}
C_l^{eq} = {\hat f^{eq}_{NL} \over (2l+1)} \sum_{l'}\sum_{l''} \left \{ {B^{eq}_{ll'l''} \hat B_{ll'l''} \over C_l C_{l'} C_{l''}} \right \}
\end{equation}
\n
and finally we can recover the one-point statistic or the cross-skewness of \citet{KSW}.
\begin{equation}
S_3^{eq} = \sum_l (2l+1) C_l^{eq} = {\hat f^{eq}_{NL}} \sum_l \sum_{l'}\sum_{l''} \left \{ {B^{eq}_{ll'l''} \hat B_{ll'l''} \over C_l C_{l'} C_{l''}} \right \}.
\end{equation}
\n
Individual contributions to the final skewness following relations, which can be useful
diagonistics for numerical checks:
\begin{equation}
\sum_l (2l+1)C_l^{A,B^2} = \sum_l(2l+1) C_l^{AB,B}; \qtwo
\sum_l (2l+1) C_l^{B,CD} = \sum_l (2l+1) C_l^{C,BD} = \sum_l (2l+1) C_l^{D,BC}.
\end{equation}
\n
Given the signal-to-noise ratio of current estimates of $S_3$ or equivalently $f_{NL}$
from WMAP surveys, it may not be possible to use many narrow bins to evaluate
the $C_l$s associated with the primordial bispectra. However,
with the increase in experimental sensitivity of future CMB experiments, it will be possible
to divide the $l$ range in narrower bins.  The important point here is that the data must show consistency with the theoretical $C_l^{loc/eq}$ to make a convincing case that the non-gaussianity is primordial.  However, these expressions are only optimal for all-sky coverage and homogeneous noise; we relax these assumptions in the next sections.


\section{Partial Sky coverage and Non-Uniform Noise: An Approximate Treatment }
\label{sec:part_sky}

It was pointed out in \citet{Babich,Crem06,Yadav08} that in the presence 
of partial sky coverage, e.g. due to the presence of a mask or because of galactic foregrounds 
and bright point sources, as well as, 
in the case of non-uniform noise, spherical symmetry is destroyed. The estimator introduced above 
will then have to be modified by adding terms linear in the observed map. The linear terms
for the local model can be written in the following form:
\begin{equation}
\hat S^{linear}_{loc} = -{1 \over f_{sky}}\int r^2dr \int d \hat \Omega  \left \{ 2B(r,\hat \Omega)\langle A(r,\hat \Omega)B(r,\hat \Omega)\rangle_{sim} 
+A(r,\hat \Omega) \langle B^2(r,\hat \Omega)\rangle_{sim} \right \}
\end{equation}
\n
The linear terms therefore are constructed from correlating the Monte-Carlo (MC) averaged $\langle A(n,r)B(n,r)\rangle_{sim}$
product maps with the input $B$ map. The mask and the noise that are used in constructing the Monte-Carlo
averaged product map are exactly same as the observed maps and the ones derived from them such as $A$ or $B$.
\begin{eqnarray}
\hat S^{linear}_{eq} = && -{18 \over f_{sky}}\int r^2dr \int d \hat \Omega \big \{ 2B(r,\hat \Omega)\langle A(r,\hat \Omega)B(\hat \Omega,r)\rangle_{sim}
+A(r,\hat \Omega) \langle B^2(r,\hat \Omega)\rangle_{sim} + 2D(r,\hat \Omega)\langle D(r,\hat \Omega)^2 \rangle \nonumber \\&& -2 B(r,\hat \Omega)\langle C(r,\hat \Omega)D(\hat \Omega ,r) \rangle_{sim} 
 -2 C(r,\hat \Omega) \langle B(r,\hat \Omega)D(r,\hat \Omega) \rangle_{sim} -2 D(r,\hat \Omega) \langle B(r,\hat \Omega)C(r,\hat \Omega) \rangle \big \}.
\end{eqnarray}
\n
Mode-mode coupling is important at low angular modes, and we consider the full case later, but for higher frequency modes, we can approximate the linear correction to the local shape:
\begin{equation}
C_l^{loc} = {1 \over f_{sky}}\left \{ {C_l^{A,B^2} - 
2C_l^{\langle A,B \rangle B} - C_l^{A, \langle B^2 \rangle}} \right \} +
{2 \over f_{sky}}\left \{ {C_l^{AB,B} - 
C_l^{\langle AB \rangle, B} -C_l^{B\langle A, B\rangle} - C_l^{A\langle B,B \rangle}} \right \}
\end{equation}
\n
where $f_{sky}$ is the sky fraction observed.

The $C_l$s such as $C_l^{\langle AB \rangle ,B}$ describe the cross-power spectra associated with
Monte-Carlo averaged product maps $\langle A(n,r)B(n,r) \rangle$ constructed with the same mask and
the noise model as the the observed map $B$. Likewise, the term  $C_l^{A\langle B,B \rangle}$
denotes the average cross-correlation computed from MC averaging, of the product map constructed
from the observed map $A(\Omega,r)$
multiplied with a MC realisation of map $B(\Omega,r)$ against the same MC realisation $B(\Omega,r)$.
\begin{eqnarray}
C_l^{eq} = && -18 \Big [ {1 \over f_{sky}} \left \{ C_l^{A,B^2} - 2C_l^{\langle A,B \rangle B} - 
C_l^{A, \langle B^2 \rangle} \right \} +
{2 \over f_{sky}}\left \{ {C_l^{AB,B} - 
C_l^{\langle AB \rangle, B} -C_l^{B\langle A, B\rangle} - C_l^{A\langle B,B \rangle}} \right \}
\nonumber \\
&&+{2 \over f_{sky}}\left \{ C_l^{D,D^2} - 2C_l^{\langle D,D \rangle D} - C_l^{D, \langle D^2 \rangle}
 \right \} - { 2 \over f_{sky}} \left \{ C_l^{B,CD} - 2C_l^{\langle B,C \rangle D} - 
C_l^{B, \langle CD \rangle} \right \} \nonumber \\
&& - { 2 \over f_{sky}} \left \{ C_l^{D,CB} - 2C_l^{\langle D,C \rangle B} - 
C_l^{D, \langle CB \rangle} \right \} 
- { 2 \over f_{sky}} \left \{ C_l^{C,BD} - 2C_l^{\langle C,B \rangle D} - 
C_l^{C, \langle BD \rangle} \right \} \Big ].
\end{eqnarray}
\n
\citet{Crem06} showed via numerical analysis that the linear terms are
less important in the equilateral case than in the local model. The use of such Monte-Carlo maps to model the effect of mask and noise greatly improves the speed compared
to full bispectrum analysis.

The use of linear terms was found to greatly reduce the scatter of the estimator, thereby improving its optimality.
The estimator was used in \citet{YaWa08} also to compute the $f_{NL}$ from combined $T$ and $E$ maps.
The analysis presented above for both the one-point statistics and the power-spectral analysis is
approximate, because it uses a crude $f_{sky}$ approximation to deconvolve the estimated power spectrum
to compare with analytical prediction. A more accurate analysis should take into account the mode-mode
coupling which can dominate at low $l$. The general expression which includes the mode-mode coupling
will be presented in the next section. However it was found out by \citet{YaWa08} that removing low
$l$s from the analysis can be efficient way to bypass the mode-mode coupling. A complete numerical
treatment for the case of two-point statistics such as $C_l^{A,B^2}$ will be presented elsewhere.

\section{Generalisation of Optimal Estimators for realistic survey strategy: Exact Analysis}
\label{sec:generalcase}

\n
The general expression for the bispectrum estimator was developed by \citet{Babich}
for arbitrary sky coverage and inhomogeneous noise. The estimator includes a cubic term,
which by matched-filtering maximises the response for a specific type of input map bispectrum.
The linear terms vanish in the absence of anisotropy but should be included for realistic noise to reduce the scatter in the estimates (see Babich 2005 for details).
We define the optimal estimator as: 
\begin{eqnarray}
&& \hat E_L[a] = \sum_{L'} [N^{-1}]_{LL'} \Big [ {1 \over 6} \sum_{MM'} \sum_{ll'l_imm'm_i} B_{L'll'}\left ( \begin{array}{ c c c }
     L' & l & l' \\
     M' & m & m'
  \end{array} \right) \nonumber \\ && ~~~~~~~~~~~\times \big \{ (C^{-1}_{L'M',l_1m_1}a_{l_1m_1})
(C^{-1}_{lm,l_2m_2}a_{l_2m_2})(C^{-1}_{l'm',l_3m_3}a_{l_3m_3}) \nonumber \\
&& ~~~~~~~~~~~~~~~~~~~~~~~~~~- C^{-1}_{lm,l'm'} (C^{-1}_{L'M',l_2m_2}a_{l_2m_2})-
 2C^{-1}_{LM,lm} (C^{-1}_{l'm',l_2m_2}a_{l_2m_2}) \big \} \Big ] 
\end{eqnarray}

\noindent
where $N_{LL'}$ is a normalization to be discussed later. A factor of ${1 /(2l+1)}$
can be introduced with the sum $\sum_M$, if we choose not to introduce
the $N_{LL'}$ normalization constant. This will make the estimator equivalent to
the one introduced in the previous section. Clearly as the data is weighted 
 by $C^{-1}= (S+N)^{-1}$, or the inverse covariance matrix,
addition of higher modes will reduce the variance of the estimator. In contrast, the performance of
sub-optimal estimators can degrade with resolution, due to the presence of inhomogeneous noise or
a galactic mask. However, a wrong noise covariance matrix
can not only make the estimator sub-optimal but it will make the estimator biased too. The noise model
will depend on the specific survey scan strategy. Numerical implementation of such inverse-variance 
weighting or multiplication of a map by $C^{-1}$ can be carried out by conjugate 
gradient inversion techniques. Taking clues from \citet{SmZa06}, we extend their estimators for the case of the bispectrum-related power spectrum. 
We will be closely following their notation whenever possible. First we define $Q_L[a]$ and its
derivative $Q_L[a]$. The required input harmonics $a_{lm}$ are denoted as $a$.
\begin{eqnarray}
&&\hat Q_L[a] \equiv \sum_{M} a_{LM} \sum_{l'm',l''m''} B_{Ll'l''}\left ( \begin{array}{ c c c }
     L & l' & l'' \\
     M & m' & m''
  \end{array} \right) a_{l'm'}a_{l''m''} \\
&& \partial_{lm} \hat Q_L[a] \equiv \delta_{Ll} \sum_{l'm',l''m''} B_{Ll'l''}\left ( \begin{array}{ c c c }
     L & l' & l'' \\
     m & m' & m''
  \end{array} \right) a_{l'm'}a_{l''m''} + 2\sum_{M} a_{LM} 
\sum_{l'm'} B_{Lll'}\left ( \begin{array}{ c c c }
     L & l & l' \\
     M & m & m'
  \end{array} \right) a_{l'm'} \nonumber \\
\end{eqnarray}
\n
These expressions differ from that of one-point estimators by the absence of an
extra summation index. $Q_L[a]$ therefore represents a map as well as  $\partial_{lm} Q_L[a]$,
however  $Q_L[a]$ is cubic in input maps $a_{lm}$ where as  $\partial_{lm} Q_L[a]$ is quadratic
in input.

The bispectrum-related power spectrum can then be written as (summation convention for the next two equations):
\begin{equation}
\hat E_L[a] = [N^{-1}]_{LL'} \left \{ Q_{L'}[C^{-1}a] - [C^{-1}a]_{lm} \langle \partial_{lm} Q_{L'}[C^{-1}a']\rangle_{MC}) \right \}
\end{equation}
\n
Here $\langle \rangle_{MC}$ denotes the Monte-Carlo averages. The inverse covariance matrix in harmonic
domain $C^{-1}_{l_1m_1,l_2m_2}= \langle a_{l_1m_1}a_{l_2m_2} \rangle^{-1}$ encodes the effects of noise and the mask. For all sky and signal-only limit,
it reduces to the usual  $C^{-1}_{l_1m_1,l_2m_2} = {1\over C_l} \delta_{ll'}\delta_{mm'}$
The normalisation of the estimator which ensures unit response can be written as:
\begin{equation}
N_{LL'} = {1 \over 3} \left \langle \left \{ \partial_{l_1m_1}Q_L[C^{-1}a] \right \} C^{-1}_{l_1m_1,l_2m_2}
\left \{ \partial_{l_2m_2}Q_{L'}[C^{-1}a]\right \} \right \rangle -{1 \over 3} \left \{ \left \langle \partial_{l_1m_1}Q_L[C^{-1}a] \right \rangle \right \} 
C^{-1}_{l_1m_1,l_2m_2} \left \{ \langle\partial_{l_2m_2}Q_{L'}[C^{-1}a]\rangle \right \}.
\end{equation}
\n
We will be using the following identity in our derivation:
\begin{equation}
\langle [C^{-1}a]_{l_1m_1} [C^{-1}a]_{l_2m_2} \rangle = C^{-1}_{l_1m_1,l_2m_2}.
\end{equation}
\n
The Fisher matrix, encapsulating the errors and covariances on the $E_L$, for a general survey associated with a specific form of bispectrum can 
finally be written as:
\begin{eqnarray}
F_{LL'}  = &&\sum_{MM'}  \sum_{l_il_i'm_im_i'}  B_{Ll_1 l_1'} B_{L'l_2 l_2'}\left ( \begin{array}{ c c c }
     L & l_1 & l_1' \\
     M & m_1 & m_1'
  \end{array} \right)\left ( \begin{array}{ c c c }
     L' & l_2 & l_2' \\
     M' & m_2 & m_2'
  \end{array} \right) \nonumber \\
&& \times {1\over 6}\big \{ 2 C^{-1}_{LM,L'M'}C^{-1}_{l_1 m_1,l_1' m_1'}C^{-1}_{l_2 m_2,l_2' m_2'} +
4C^{-1}_{LM,L'M'}C^{-1}_{l_1 m_1,l_1' m_1'}C^{-1}_{l_2 m_2,l_2' m_2'}\big \} = 
{1 \over 36} \left \{ 2\alpha^{PP}_{LL'} + 4\alpha^{QQ}_{LL'} \right \}.
\end{eqnarray}
\n
Using the following expressions which are extension of \citet{SmZa06},
we find that the Fisher matrix can be written as sum of two $\alpha$
terms $\alpha^{PP}$ and $\alpha^{QQ}$. The alpha terms correspond to coupling only of modes 
that appear in different $3j$ symbols. Self couplings are represented by the beta terms. 
The subscripts describes the coupling of various $l$ and $L$
indices. The subscript $PP$ correspond to coupling of free indices,i.e one free index $L_1$ 
with another free index $L_2$ and similar coupling for indices that are summed over such as $l_1$,
$l_2$ etc. Similarly for subscript $QQ$ the free indices are coupled with summed indices.
Couplings are represented by the inverse covariance matrices in harmonic domain e.g. $C^{-1}_{lm,LM}$
denotes coupling of mode $LM$ with $lm$.
\begin{eqnarray}
&& \alpha^{PP}_{L_1L_2} =\sum_{M_1,M_2} \sum_{l_i l_i' m_i m_i'} B_{ L_1l_1l_1'} B_{L_2l_2l_2'}
\left ( \begin{array}{ c c c }
      L_1 & l_1 & l_1' \\
     M_1 & m_1 & m_1'
  \end{array} \right)\left ( \begin{array}{ c c c }
      L_2 & l_2 & l_2' \\
     M_2 & m_2 & m_2'
  \end{array} \right)  
C^{-1}_{L_1M_1,L_2M_2}C^{-1}_{l_1m_1,l_2m_2}C^{-1}_{l_1'm_1',l_2'm_2'}\\
&& \alpha^{QQ}_{L_1L_2} =\sum_{M_1,M_2} \sum_{l_il_i'm_im_i'}
 B_{L_1l_1l_1'} B_{L_2l_2l_2'}
\left ( \begin{array}{ c c c }
     L_1 & l_1 & l_1' \\
     M_1 & m_1 & m_1'
  \end{array} \right)\left ( \begin{array}{ c c c }
     L_2 & l_2 & l_2' \\
     M _2& m_2 & m_2'
  \end{array} \right)  C^{-1}_{L_1M_1,l_2m_2}C^{-1}_{l_1m_1,L_2M_2}C^{-1}_{l_1'm_1',l_2'm_2'}\\
&& \alpha^{PP}_{L_1L_2} = \wick[d]{123}{(<1 L_1 <2 l_1 <3 l_1')(>1 L_2 >2 l_2 >3 l_2')};~~~~~
\alpha^{QQ}_{L_1L_2} = \wick[d]{123}{(<1 L_1 <2 l_1 <3 l_1')(>2 L_2 >1 l_2 >3 l_2')}.
\end{eqnarray}
\n
These results will reduce to those of \citet{SmZa06} when further summations 
over $L_1$ and $L_2$ are introduced to collapse the two-point object to the corresponding
one-point quantity. The beta terms that denote cross-coupling can be written as:
\begin{eqnarray}
&& \beta^{PP}_{L_1L_2} =\sum_{M_1,M_2} \sum_{l_il_i'm_im_i'}
 B_{L_1l_1l_1'} B_{L_2l_2l_2'}
\left ( \begin{array}{ c c c }
     L_1 & l_1 & l_1' \\
     M_1 & m_1 & m_1'
  \end{array} \right)\left ( \begin{array}{ c c c }
     L_2 & l_2 & l_2' \\
     M_2 & m_2 & m_2'
  \end{array} \right) C^{-1}_{L_1M_1,L_2M_2}C^{-1}_{l_1m_1,l_1'm_1'}C^{-1}_{l_2m_2,l_2'm_2'} \\
&& \beta^{PQ}_{L_1L_2} =\sum_{M_1,M_2} \sum_{l_i l_i' m_i m_i'}
 B_{L_1l_1 l_1'} B_{L_2 l_2 l_2'}
\left ( \begin{array}{ c c c }
     L_1 & l_1 & l_1' \\
     M_1 & m_1 & m_1'
  \end{array} \right)\left ( \begin{array}{ c c c }
     L_2 & l_2 & l_2' \\
     M_2 & m_2 & m_2'
  \end{array} \right) C^{-1}_{L_1M_1,l_2 m_2}C^{-1}_{l_1 m_1,l_1'm_1'}C^{-1}_{L_2 M_2,l_2'm_2'} \\
&& \beta^{QQ}_{L_1L_2} =\sum_{M_1,M_2}\sum_{(l_il_i'm_im_i'}
 B_{L_1l_1l_1'} B_{L_2l_2 l_2'}
\left ( \begin{array}{ c c c }
     L_1 & l_1 & l_1' \\
     M_1 & m_1 & m_1'
  \end{array} \right)\left ( \begin{array}{ c c c }
     L_2 & l_2 & l_2' \\
     M_2 & m_2 & m_2'
  \end{array} \right)  C^{-1}_{L_1M_1,l_1m_1}C^{-1}_{L_2M_2,l_2m_2}C^{-1}_{l_1'm_1',l_2'm_2'}\\
&& \beta^{PP}_{L_1L_2} = \wick[u]{123}{(<1 L_1 <2 l_1 >2 l_1')(>1 L_2 <3 l_2 >3 l_2')};~~~~~
\beta^{PQ}_{L_1L_2} = \wick[u]{123}{(<1 L_1 <2 l_1 >2 l_1')(<3 L_2 >1 l_2 >3 l_2')};~~~~~ 
\beta^{QQ}_{L_1L_2} = \wick[u]{123}{(<1 L_1 >1 l_1 <3 l_1')(<2 L_2 >2 l_2 >3 l_2')}.
\end{eqnarray}
\n
No summation over repeated indices is assumed. Using these expressions one can finally show that
\begin{equation}
 [F^{-1}]_{LL'} =\langle \hat E_L \hat E_{L'} \rangle - \langle \hat E_L \rangle \langle \hat E_{L'} \rangle  =  {\langle AA + BB + CC +2AB +2BC +2AC} \rangle_{LL'}
\end{equation}
where
\begin{eqnarray}
&& AA_{LL'}= \left \{ {2\over 36} \alpha^{PP}_{LL} + {4\over 36} \alpha^{PP}_{LL'}  +
{2\over 36} \beta^{PP}_{LL} + {4\over 36} \beta^{PP}_{LL'}+ {4\over 36} \beta^{PP}_{LL'}   \right \};
~~~~~~~~  BB_{LL'}= \beta^{PP}_{LL'};
~~~~~~~~  CC_{LL'}= 4\beta^{QQ}_{LL'}\\
&& 2AB_{LL'} = -2(\beta^{PP}_{LL}+2\beta^{QP}_{LL});  
~~~~~~~~ 2AC_{LL'} = -4(2\beta^{QQ}_{LL'} + \beta^{PQ}_{LL'});
~~~~~~~~  2BC_{LL'} = 4\beta^{PQ}_{LL'}.
\end{eqnarray}
\n
The final expression can be written in terms of only $\alpha$ terms as the $\beta$ terms cancel out:
\begin{equation}
F_{LL'} = \left \{ {2\over 36} \alpha^{PP}_{LL'} + {4\over 36} \alpha^{QQ}_{LL'} \right \}.
\end{equation}
\n
If we sum over $LL'$ the Fisher matrix reduces to a scalar $F=\sum_{LL'}F_{LL'}$ 
with, $\alpha^{PP}_{LL'} = \alpha^{QQ}_{LL'}= \alpha$ and   $\beta^{PP}_{LL'} = \beta^{PQ}_{LL'} =\beta^{QQ}_{LL'}= \beta$, where $\alpha$, $\beta$ and $F$ are exactly the same as introduced in \citet{SmZa06}.

\subsection{Joint Estimation of Multiple bispectrum-related Power-Spectra}

\n
The estimation technique described above can be generalised to take cover the bispectrum-related power spectrum associated with different set of bispectra (X,Y):
\begin{equation}
\hat E_L[a] = [F^{-1}]_{LL'}^{XY} \left \{ Q_{L'}^Y[C^{-1}a] - [C^{-1}a]_{lm} \langle \partial_{lm} Q_{L'}^Y[C^{-1}a]\rangle_{MC}) \right \}.
\end{equation}
\n
The associated Fisher matrix now will consist of sectors $F_{LL'}^{XX}$,$F_{LL'}^{YY}$ and  $F_{LL'}^{XY}$.
The sector $XX$ and $YY$ will in general will be related to errors associated with estimation of 
bispectra of $X$ and $Y$ types, whereas the sector $XY$ will correspond to their cross-correlation.
\begin{equation}
F_{LL'}^{XY}  = \left \{ {2\over 36} \left [ \alpha^{PP}_{LL'} \right ]^{XY} + {
4\over 36} [\alpha^{QQ}_{LL'}]^{XY} \right \} 
\end{equation}
\n
where we have:
\begin{equation}
\left[\alpha^{PP}_{LL'} \right ]^{XY} =
\sum_{MM'}
 \sum_{l_i l_i' m_i m_i'} B^X_{Ll_1l_1'} B^Y_{L'l_2l_2'} \left ( \begin{array}{ c c c }
     L & l_1 & l_1' \\
     M & m_1 & m_1'
  \end{array} \right)\left ( \begin{array}{ c c c }
     L' & l_2 & l_2' \\
     M' & m_2 & m_2'
  \end{array} \right)  C^{-1}_{LM,L'M'}C^{-1}_{l_1m_1,l_2m_2}C^{-1}_{l_1'm_1',l_2'm_2'}\\
\end{equation}
\n
and a similar expression holds for $[\alpha^{QQ}_{LL'}]^{XY}$.

\subsection{Generalisation to non-optimal weights}

Although the estimator as constructed is fully optimal - its true usefulness is
determined by the affordability of the construction of the $C^{-1}$ matrix as well as the availability
of a fast method to multiply it with CMB maps in a Monte Carlo chain.
A more general class of estimator which is sub-optimal can be constructed by replacing
the inverse covariance weighting of the data $[C^{-1}a]$ by $[R a]$, where $[R]$ is
an arbitrary filter function. In this case the estimator with unit response can be written
as:
\begin{equation}
\hat E^R_L[a] = \sum_{L'} \left [ F^{-1} \right ]_{LL'}\left \{ Q_{L'}[Ra] -  [Ra]_{lm} \left \langle \partial_{lm} Q_{L'}[Ra]\right \rangle_{MC} \right \}
\end{equation} 
\n
where the normalisation and the variance of the estimator can be constructed in a similar manner:
\begin{eqnarray}
 F_{LL'}& =& \langle(\hat E_L)(\hat E_{L'})\rangle - \langle \hat E_L \rangle \langle \hat E_{L'} \rangle^{-1} = { 1 \over 3 } \left \langle\left \{ \partial_{lm}Q_L[Rs] \right \} R \left \{\partial_{lm}Q_{L'}[Rs]\right \}  \right \rangle_{MC}\nonumber\\
%
&=&{1 \over 3} \langle \left \{ \partial_{l_1m_1}Q_l[Ra] \right \} [FCF]_{l_1m_1,l_2m_2}
\left \{ \partial_{l_1m_1}Q_{l'}[Ra]\right \} \rangle -{1 \over 3} \left \{ \langle \partial_{l_1m_1}Q_l[Ra] \rangle \right \} 
C^{-1}_{l_1m_1,l_2m_2} \left \{ \langle\partial_{l_2m_2}Q_{l'}[Ra]\rangle \right \}.
\end{eqnarray}
\n
The optimal weighting
can be replaced by an arbitrary weight or no weighting at all (R=I). However in this case the
estimator though unbiased clearly becomes a sub-optimal one.

\subsection{Recovery of all-sky homogeneous noise model}

In the the all-sky limit we recover the usual expression:
%
\begin{equation}
F_{LL'} = {1 \over 36}  \left \{ 2\sum_{ll'}{{B_{Lll'}^2 \over C_l C_{L} C_{L'}} \delta_{LL'} +  
4\sum_{l} {B_{LL'l}^2 \over C_l C_{L} C_{L'}}} \right \}.
\end{equation}
\n
In the case of joint analysis as before we can write down the off-diagonal blocks 
of the Fisher matrix as:
\begin{equation}
F_{LL'}^{XY} = {1 \over 36}  \left \{ 2\sum_{ll'}{{{B_{Lll'}^X B_{Lll'}^Y \over C_{L} C_{l} C_{l'}}} \delta_{LL'} +  
4\sum_{l} {B_{LL'l}^XB_{LL'l}^Y \over C_{L} C_{L'} C_{l}}} \right \}.
\end{equation}
\n
For $X=Y$ we recover the diagonal blocks of the Fisher matrix
for the independent estimations derived before. The errors for independent
estimates are given by $\sqrt{(F_{LL}^{XX})^{-1}}$,
where as for the joint estimation the errors are  $\sqrt{(F_{LL}^{YY})^{-1}}$.

\subsection{More general bispectra}

A more general bispectrum can be written as a sum of individual product terms:
\begin{equation}
b_{l_1l_2l_3} = {1 \over 6}\sum_{i}^{N_{fact}} A_{l_1}^{i} B_{l_2}^{i} C_{l_3}^{i} + {\rm symm.perm.}
\end{equation}
\n
This can be seen as a generalization of the type of bispectra introduced for the
equilateral case. It may also possible to approximate the bispectrum  
$b_{l_1l_2l_3}$ to a smaller number of optimum factorizable terms $N_{opt}$ 
with suitable weight factors $w_i$
as given in \citet{SmZa06}. In this case the bispectrum is expressed as:
\begin{equation}
b_{l_1l_2l_3} = {1 \over 6}\sum_{i}^{N_{opt}}w_i A_{l_1}^{i} B_{l_2}^{i} C_{l_3}^{i} + {\rm symm.perm.}
\end{equation}
\n
Clearly significant computational gain can only be achieved if $N_{opt} \ll N_{fact}$. In the following discussion we generalise the description in previous sections
to such composite bispectra. Following the same analytical reasoning we can show that
the Fisher matrix elements for such a composite bispectrum can be written as:
\begin{eqnarray}
\!\! [F_{LL'}^{XY}]_{ij} =
{1 \over 36}  \Big \{  2\delta_{LL'} \sum_{ll'}  {(2L+1)(2l+1)(2l'+1) \over 144\pi}  
\left ( \begin{array}{ c c c }
     L & l & l' \\
     0 & 0 & 0
  \end{array} \right)^2 {1 \over C_L C_{l} C_{l'}}
\big [  A_{L}^{i} B_{l}^{i} C_{l'}^{i} + \dots \big ]^X \big [
[  A_{L}^{j} B_{l}^{j} C_{l'}^{j} + \dots \big ]^Y \nonumber\\
+ 4\sum_l {(2L+1)(2L'+1)(2l+1) \over 144\pi}  
\left ( \begin{array}{ c c c }
     L & L' & l \\
     0 & 0 & 0
  \end{array} \right)^2{1 \over C_L C_{L'} C_{l}}
\big [  A_{L}^{i} B_{L'}^{i} C_{l}^{i} + \dots \big ]^X \big [
[ A_{L}^{j} B_{L'}^{j} C_{l}^{j} + \dots \big ]^Y 
\Big \}.
\end{eqnarray}
\n
The symbols $A^{i}$, $B^{i}$ etc denotes the $i$-th  term in the factorised representation
of a specific type of the bispectrum of type $X$ or $Y$. The total contribution of all terms will constitute the final Fisher matrix:
\begin{equation}
F_{LL'}^{XY} = \sum_{ij} [F_{LL'}^{XY}]_{ij}; ~~~~~~ F^{XY} = \sum_{LL'}F_{LL'}^{XY}.
\end{equation}
\n 
Using the following identity we can project this expression onto real-space:
\begin{equation}
\int_{-1}^{1} dz  P_{l_1}(z)P_{l_2}(z)P_{l_3}(z) = 
2 \left ( \begin{array}{ c c c }
     l_1 & l_2 & l_3 \\
     0 & 0 & 0
  \end{array} \right)^2
\end{equation}
\begin{eqnarray}
[F_{LL'}]_{ij}=  \int_{-1}^{1} dz \left [ \left ( \xi^{A^iA^j}_L(z) \xi^{B^iB^j}(z) \xi^{C^iC^j}(z) + \dots \right )\delta_{LL'}
 + \left ( \xi^{A^iA^j}_{L}(z) \xi^{B^iB^j}_{L'}(z) \xi^{C^iC^j}(z) + \dots \right ) \right ]
\end{eqnarray}
\n
The first term describes the diagonal entries of the Fisher matrix and the second term relates to
the off-diagonal terms. 
\begin{eqnarray}
\xi^{A^iA^j}(z)= \sum_l \xi^{A^iA^j}_{l}(z);\quad{\rm where}\quad
\xi^{A^iA^j}_l (z) \equiv  {2l+1 \over 4 \pi} {A^i_l A^j_l  \over C_l} P_l(z).
\end{eqnarray}
\n
In case of cross-correlational studies the $A^i$ and $A^j$ will come from two different factorization of distinct bispectra denoted before by $X$ and $Y$. The case of weighted sum can also
be derived in an exactly similar manner.

\section{Models for Non-Gaussianity}

We will use two commonly-used specific models for the primordial non-Gaussianity as well as 
one foreground source of contamination i.e. extra-galactic point sources to demonstrate the 
power of our statistics in this section.

\subsection{Local or squeezed Model}

Using the specific form for $b^{loc}_{l_1l_2l_3}$ the Fisher matrix elements for the local model can be expressed in terms of the functions $\alpha$ and $\beta$ and the power spectrum
$C_{l}$.
\begin{eqnarray}
\alpha_{L_1L_2}^{PP} = && {f_{NL}^2 \over 4\pi} (2L_1+1)(2L_2+1)\sum_{l} \left ( \begin{array}{ c c c }
     L_1 & L_2 & l \\
     0 & 0 & 0
  \end{array} \right)^2 { 1 \over C_{L_1}C_{L_2}C_l} \nonumber \\
&& \times \left \{ \int r^2 dr \left ( 2\alpha_{L_1}(r)\beta_{L_2}(r)\beta_l(r) +  \alpha_{l}(r)\beta_{L_1}(r)\beta_{L_2}(r) \right )\right \}^2  
  \\
\alpha_{L_1L_2}^{QQ} = && \delta_{L_1L_2}~~ {f_{NL}^2 \over 4\pi}(2L_1+1) \sum_{l_1l_2} (2l_1+1)(2l_2+1)\left ( \begin{array}{ c c c }
     L_1 & l_1 & l_2 \\
     0 & 0 & 0
  \end{array} \right)^2 { 1 \over C_{L_1}C_{l_1}C_{l_2}} \nonumber \\ && \times  
\left \{\int r^2 dr \Big ( \alpha_{L_1}(r)\beta_{l_1}(r)\beta_{l_2}(r) +  \alpha_{l_1}(r)\beta_{l_2}(r)\beta_{L_1}(r) + \alpha_{l_2}(r)\beta_{l_1}(r)\beta_{L_1}(r) \Big ) \right \}^2.
\end{eqnarray}

\subsection{Equilateral Model}

\n
Using the equilateral model the contribution to the Fisher 
matrix can be expressed as:
\begin{eqnarray}
\alpha_{L_1L_2}^{PP} = && {f_{NL}^2 \over 4\pi}(2L_1+1)(2L_2+1)
\sum_{l}(2l+1) \left ( \begin{array}{ c c c }
     L_1 & l_1 & l_2 \\
     0 & 0 & 0
  \end{array} \right)^2 { 1 \over C_{L_1}C_{L_2}C_{l}} \nonumber \\
&& \times \left \{ \int r^2dr \left ( -\alpha_{L_1}(r) \beta_{L_2}(r) \beta_{l}(r) 
+ \delta_{L_1}(r)\delta_{L_2}(r)\delta_{l}(r) 
+ \beta_{L_1}(r)\gamma_{L_2}(r)\delta_{l}(r)+ {\rm cyc.perm.}
\right )\right \}^2.\\
\alpha_{L_1L_2}^{QQ} = && \delta_{L_1L_2}(2L_1+1) {f_{NL}^2 \over 4\pi} 
\sum_{l_1l_2}(2l_1+1)(2l_2+1)\left ( \begin{array}{ c c c }
     L_1 & l_1 & l_2 \\
     0 & 0 & 0
  \end{array} \right)^2 { 1 \over C_{L_1}C_{l_1}C_{l_2}} \nonumber \\
&&\times \left \{ \int r^2dr \left ( -\alpha_{L_1}(r) \beta_{l_1}(r) \beta_{l_2}(r) 
+ \delta_{L_1}(r)\delta_{l_2}(r)\delta_{l_3}(r) 
+ \beta_{L_1}(r) \gamma_{l_1}(r)\gamma_{l_2}(r)+ {\rm cyc.perm.}
\right ) \right \}^2
\end{eqnarray}
\n
The power spectra $C_l$ appearing in the denominator take contributions both
from the pure signal or CMB and the detector noise. It is possible to bin 
the estimates in large enough bins to report uncorrelated estimates
which may be possible for an experiment such as Planck with very high sky-coverage. A detailed analysis of the singularity structure of the error-covariance
matrix will be presented elsewhere.

\subsection{Point Sources}

\n
The bispectrum from residual point-sources which are assumed random can
be modelled as $b_{l_1l_2l_3}=b_{ps}$. The exact value depends on
the flux limit and the mask used in the survey. The accuracy of such 
an approximation can indeed be extended by adding contributions from
correlation terms.
\begin{eqnarray}
&& \alpha_{L_1L_2}^{PP} = {b_{ps}^2\over 4\pi} (2L_1+1)(2L_2+1)\sum_{l} 
 \left ( \begin{array}{ c c c }
     L_1 & L_2 & l \\
     0 & 0 & 0
  \end{array} \right)^2  { 1 \over C_{L_1}C_{L_2}C_{l}} \\
&& \alpha_{L_1L_2}^{QQ} = \delta_{L_1L_2} {b_{ps}^2\over 4\pi} (2L_1+1)\sum_{l_1,l_2} 
 \left ( \begin{array}{ c c c }
     L_1 & l_1 & l_2 \\
     0 & 0 & 0
  \end{array} \right)^2 { 1 \over C_{L_1}C_{l_1}C_{l_2}}
\end{eqnarray}
\n
Similar computations can be done for cross-correlation among various
contributions, e.g. contamination due to point sources or estimation
of a specific type of non-Gaussianity. A joint estimation is useful for finding out 
also the level of cross-contamination from one theoretical model while another is being estimated.

\section{Conclusions}
\label{sec:conclusion}

We have addressed the problem of finding an estimator of primordial non-Gaussianity from microwave background data.  The new feature of this analysis is that the technique presented here allows one to make an assessment of whether any non-Gaussian signal is primordial or not.  The issue here is that if one finds an estimate of a level of primordial non-Gaussianity which is inconsistent with zero, then it is very difficult at present to make a convincing case that it is indeed primordial and not simply contamination by any number of other effects which might lead to a non-Gaussian CMB map.  The method does this by performing a less aggressive data compression than previous analyses.  Rather than compressing the data to a single number (typically an estimate of $f_{NL}$), it reduces the data to a function, the {\em bispectrum-related power spectrum}.  This is an average cross-power spectrum of certain maps constructed from the CMB data. By doing this construction, one retains the ability to assess the contributions from different sources, such as residual point sources, incomplete foreground subtraction and so on (see e.g. \citet{SerCoo08}).  As an example, we have computed the expected bispectrum-related power spectrum optimised for local non-gaussianity (we also consider the equilateral type), and calculated the contribution expected from an unclustered population of point sources.   Indeed, one can use standard statistical methods to estimate the amplitude of components of non-Gaussianity, and since these contributors have quite different harmonic dependences, the estimators will be largely decoupled.  The power of the technique will depend on the level of the primordial signal, but if it is at the level claimed by \citet{YaWa08}, then it will be possible with Planck data to construct a large number of band-power estimates of the bispectrum-related power spectrum to see if it is primordial.    We also include polarisation in the analysis (see appendix). The work draws extensively on previous studies, in particular generalising the \citet{KSW} analysis for the all-sky, homogeneous noise case.  For the more realistic case of partial sky coverage and inhomogeneous noise, we present optimised estimators of the bispectrum-related power spectrum, including linear terms and extending the work of \citet{Babich,SmZa06,SmSeZa09}.     For studies such as this, it is normal to assume the background cosmology is known from the power spectrum, but uncertainties will propagate into the $f_{NL}$ estimates \citep{LigRio08} and will be investigated elsewhere.   Note that the techniques here could be generalised to higher-order statistics such as the trispectrum, should the bispectrum vanish for some symmetry reason.

\section*{Acknowledgements}
DM acknowledges financial support from an STFC rolling grant at the University of Edinburgh. 
DM acknowledges useful discussions with Michele Liguori, Patrick Valageas 
and Anthony Lewis at various stages of the work. Use of the CMBFAST package for calculating 
the transfer function is acknowledged.

\appendix

\section{Joint Analysis of Temperature and Polarization}

\n 
Most current constraints on non-Gaussianity still come from temperature maps, but with WMAP and Planck, the situation is changing.   Similar calculations can be performed for joint temperature and E-type polarisation analysis, and the estimators 
discussed above can be generalised to include $E$ -type polarisation to tighten
the constraints. The functions $\alpha_l$ and $\beta_l$ that we discussed in the main
text needs to be generalised for both $T$, temperature and $E$-type polarization.
We follow the discussion in \citet{YKW}, see also \citet{BaZa04,Liguori07} 
for related discussions.
\begin{eqnarray}
 \alpha_l^X(r) \equiv {2 \over \pi} \int_0^{\infty} k^2 dk P_{\Phi}(k)\Delta_l^X(k) j_l(kr) ~~~~~~
 \beta_l^X(r) \equiv {2 \over \pi} \int_0^{\infty} k^2 dk \Delta_l^X(k) j_l(kr) .
\end{eqnarray}
\n
The index $X$ can be $T$ or $E$. Similar calculations can in principle be performed for
the equilateral case. We will focus however here only on the local or squeezed model. The power spectra now can be a temperature $(TT)$-only power spectrum or 
an Electric-Electric $(EE)$ polarization spectra which we define below, with specific
choice of  $\Delta_l$.
\begin{equation}
C_l^{XY}(r) \equiv {2 \over \pi} \int_0^{\infty} k^2 dk P_{\Phi}(k)\Delta_l^X(k)\Delta_l^Y(k).
\end{equation}
\n
We arrange the all-sky covariance matrix (in the harmonic domain) in a matrix, which 
takes the following form in terms of the $C_l$ defined above:
\begin{eqnarray}
[{\mathit C}]_l = \left [ \begin{array}{ c c }
     C_l^{TT} & C_l^{TE}  \\
     C_l^{TE} & C_l^{EE}
  \end{array} \right ] .
\end{eqnarray}
\n
We can now construct the $A(r,\hat \Omega)$ and $B(r,\hat \Omega)$ fields now
can be constructed using the full covariance matrix instead of only temperature
data.
\begin{equation}
A(r,\hat \Omega) = \sum_{lm} \sum_{ip} \left [ C^{-1} \right ]^{ip}_{l} 
a_{lm}^i \alpha_l^p(r)Y_{lm}(\hat \Omega)
\end{equation}
\begin{equation}
B(r,\hat \Omega) = \sum_{lm} \sum_{ip} \left [ C^{-1} \right ]^{ip}_{l} 
a_{lm}^i \beta_l^p(r)Y_{lm}(\hat \Omega).
\end{equation}
\n
The construction now follows exactly the same steps as depicted for the temperature 
case. We compute the cross-correlation of $B^2(r,\hat \Omega)$ with the $A(r,\hat \Omega)$.
The  $B^2(r,\hat \Omega)$ can be decomposed in terms of the $T$ and $E$ harmonics 
both $a_{lm}^j$ ($j$ here takes values $T$ or $E$).
\begin{eqnarray}
&& B_{lm}^{(2)}(r) = \int d\Omega B^2(r,\Omega)Y_{lm}(\Omega) \nonumber \\
&=& \sum_{l'm'} \sum_{l''m''} {\beta_{l'}(r)\left [ C^{-1} \right ]^{jq}_{l'}}
{\beta_{l''}(r)\left [ C^{-1} \right ]^{kr}_{l''} }
 \sqrt {(2l+1)(2l'+1)(2l''+1) \over 4\pi}\left ( \begin{array}{ c c c }
     l & l' & l'' \\
     0 & 0 & 0
  \end{array} \right)
\left ( \begin{array}{ c c c }
     l & l' & l'' \\
     m & m' & m''
  \end{array} \right)a_{l'm'}^j a_{l''m''}^k.
\end{eqnarray}
\n
The cross-correlation of the associated $A$ and $B$ field will contain information
both from temperature and polarization maps.
\begin{eqnarray}
&&  C_l^{A,B^2} = \int r^2 dr C_l^{A,B^2}(r) = \int r^2 dr B_{lm}^2(r) A_{lm}(r)  
=  {1 \over 2l+1} \sum_m \sum_{l'm'} \sum_{l''m''} \int r^2 dr {\alpha_{l}(r)\left [ C^{-1} \right ]^{ip}_{l}}
 {\beta_{l'}(r)\left [ C^{-1} \right ]^{jq}_{l'}}
{\beta_{l''}(r)\left [ C^{-1} \right ]^{kr}_{l''} } \nonumber \\
&& ~~~~~~~~~~~~~~~~~~~~~~~~\times \sqrt {(2l+1)(2l'+1)(2l''+1) \over 4\pi}\left ( \begin{array}{ c c c }
     l & l' & l'' \\
     0 & 0 & 0
  \end{array} \right)
\left ( \begin{array}{ c c c }
     l & l' & l'' \\
     m & m' & m''
  \end{array} \right)a_{lm}^i a_{l'm'}^j a_{l''m''}^k
\end{eqnarray}
\begin{equation}
B^{ijk}_{ll'l''} = \sum_{mm'm''} \left ( \begin{array}{ c c c }
     l & l' & l'' \\
     m & m' & m''
  \end{array} \right) a_{lm}^ia_{l'm'}^ja_{l''m''}^k
\end{equation}
\n
The mixed-bispectrum $B^{pqr}_{ll'l''}$, contains information about three-point
correlation in harmonic space with the possibility that the harmonics $a_{lm}^i$
can either be of $T$ or temperature type or $E$ or electric-polarisation type. 
The theoretical model for such bispectrum depends on functions $\alpha_l^p(r)$
and $\beta^q(r)$ where again depending on the superscript the functions can 
be of temperature or the electric type.
\begin{equation}
B^{pqr}_{ll'l''} =
\sqrt {(2l+1)(2l'+1)(2l''+1) \over 4\pi}\left ( \begin{array}{ c c c }
     l & l' & l'' \\
     0 & 0 & 0
  \end{array} \right)
\int r^2 dr \left \{\beta_{l}^p(r)\beta_{l'}^q(r)\alpha_{l''}^r(r) + \beta_{l}^p(r)\alpha_{l'}^q(r)\beta_{l''}^r(r) + 
\alpha_{l}^p(r)\beta_{l'}^q(r)\beta_{l''}^r(r) \right \}.
\end{equation}
\n
Finally in an analogous way we can express the power-spectrum related to the mixed-bispectrum
as follows:
\begin{equation}
(2l+1)(C_l^{A,B^2}+2C_l^{AB,B})= {\hat f^{NL}} \sum_{l'} \sum_{l''} \left \{ B_{ll'l''}^{ijk}
\left [ C^{-1} \right ]^{ip}_{l} 
\left [ C^{-1} \right ]^{jq}_{l'} 
\left [ C^{-1} \right ]^{kr}_{l''}
B_{ll'l''}^{pqr} \right \}.
\end{equation}
\n
We have assumed summation of repeated indices which denote the polarisation types, i.e.
${i,j,k}$ and ${p,q,r}$ in the above expression. The one-point mixed skewness
which is related to the above power spectrum is analogously written as follows:
\begin{equation}
S^{prim}_{TE} = \sum_l (2l+1)(C_l^{A,B^2}+2C_l^{AB,B})= {\hat f^{NL}} \sum_l \sum_{l'} \sum_{l''} \left \{ B_{ll'l''}^{ijk}
\left [ C^{-1} \right ]^{ip}_{l} 
\left [ C^{-1} \right ]^{jq}_{l'} 
\left [ C^{-1} \right ]^{kr}_{l''}
B_{ll'l''}^{pqr} \right \}.
\end{equation}
\n
This generalises the temperature-only power spectrum estimator introduced in the text of this paper and can be extended to take into account other models, sky-coverage and secondary anisotropies in an analogous manner.

\end{document}